\documentclass[twocolumn]{IEEEtran}

\usepackage{amsmath, graphics,amssymb,epsfig,subfigure}
\usepackage{algorithm}
\usepackage{algorithmic}
\usepackage{float}
\usepackage{multirow}
\usepackage{cite}

\newlength{\capwidth}
\newtheorem{Theorem}{Theorem}

\newtheorem{Lemma}{Lemma}

\newcommand{\yv}{\mbox{$\bf y $}}
\newcommand{\av}{\mbox{$\bf a $}}
\newcommand{\bv}{\mbox{$\bf b $}}

\newcommand{\Yv}{\mbox{$\bf Y $}}
\newcommand{\Hv}{\mbox{$\bf H $}}

\newcommand{\Av}{\mbox{$\bf A $}}
\newcommand{\xv}{\mbox{$\bf x $}}

\newcommand{\Xv}{\mbox{$\bf X $}}

\newcommand{\nv}{\mbox{$\bf n $}}

\newcommand{\ev}{\mbox{$\bf e $}}
\newcommand{\Bv}{\mbox{$\bf B $}}
\newcommand{\Qv}{\mbox{$\bf Q $}}

\newcommand{\Wv}{\mbox{$\bf W $}}

\newcommand{\Iv}{\mbox{$\bf I $}}
\newcommand{\Rv}{\mbox{$\bf R $}}

\newcommand{\Lv}{\mbox{$\bf L $}}
\newcommand{\Gv}{\mbox{$\bf G $}}

\newcommand{\Uv}{\mbox{$\bf U $}}
\newcommand{\Vv}{\mbox{$\bf V $}}

\newcommand{\be}{\begin{equation}}

\newcommand{\ee}{\end{equation}}
\newcommand{\bea}{\begin{eqnarray}}
\newcommand{\eea}{\end{eqnarray}}
\newcommand{\bdp}{\begin{displaymath}}
\newcommand{\edp}{\end{displaymath}}

\begin{document}
\title{\huge{Achievable Diversity-Rate Tradeoff of MIMO AF Relaying Systems with MMSE Transceivers}}

\author{\authorblockN{Changick Song and Cong Ling} \\
\thanks{The authors are with the Department of Electrical and Electronic Engineering, Imperial College, London, UK
(e-mail: \{csong and c.ling\}@imperial.ac.uk).}
}\maketitle

\vspace{-50pt}
\begin{abstract}
This paper investigates the diversity order of
the minimum mean squared error (MMSE) based optimal transceivers in multiple-input multiple-output (MIMO) amplify-and-forward (AF) relaying systems.
While the diversity-multiplexing tradeoff (DMT) analysis accurately predicts the behavior of the MMSE receiver for the positive multiplexing gain,
it turned out that the performance is very unpredictable via DMT for the case of fixed rates,
because MMSE strategies exhibit a complicated rate dependent behavior.
In this paper, we establish the diversity-rate tradeoff performance of MIMO AF relaying systems with the MMSE transceivers as a closed-form for all fixed rates,
thereby providing a complete characterization of the diversity order together with the earlier work on DMT.
\end{abstract}

\vspace{-0pt}
\section{Introduction}\label{sec:introduction}

In multiple-input multiple-output (MIMO) systems, although suboptimal, the linear minimum mean squared error (MMSE)
receivers have widely been adopted as a low complexity alternative to the optimal maximum likelihood (ML) receiver.
This leads to a large amount of research on the performance of MMSE receivers \cite{Hedayat:07,Kumar:09,Mehana:12},
but their performance is not fully understood yet in MIMO relaying channels.

A fundamental criterion to evaluate the performance of a MIMO system is the {\it ``diversity-multiplexing tradeoff'' (DMT)}.
Thus, many analyses have been conducted based on the DMT in MIMO relaying systems \cite{DGunduz:10,Oleveque:10,Changick:12JSAC}.
Under the MMSE strategy, however, the DMT is not sufficient to characterize the diversity order, because
the DMT framework an asymptotic notion for the high signal-to-noise ratio (SNR),
cannot distinguish between different spectral efficiencies that correspond to the same multiplexing gain which we denote by $r$.

In fact, it is known in point-to-point (P2P) MIMO channels
that while the DMT analysis accurately predicts the behavior of the MMSE receiver for the positive multiplexing gain ($r>0$),
the extrapolation of the DMT to $r=0$ is unable to predict the performance especially at low rates.
This rate-dependent behavior of MMSE receivers has first been observed by Hedayat in \cite{Hedayat:07} and comprehensively analyzed by Mehana in \cite{Mehana:12}
by performing the {\it ``diversity-rate tradeoff (DRT)''} analysis for all fixed rates.
A similar phenomenon can be observed in MMSE-based MIMO AF relaying systems, but the analysis has not been made so far.


In this paper, we investigate the achievable DRT of the linear MMSE transceivers in MIMO amplify and forward (AF) relaying systems
for all fixed data rates, where the relay transceiver and the destination receiver are jointly optimized with respect to the MMSE.
The optimal MMSE transceiver designs have been proposed in \cite{Guan:08} and \cite{Changick:09TWC} using different approaches.
In this paper, we focus on the method in \cite{Changick:09TWC} based on the error covariance decomposition,
which allows further analysis tractable.
In fact, the DRT analysis does not impose any restriction on the number of antennas at each node,
because a certain diversity gain is always achievable at arbitrarily low rates.
Thus, we first provide a new result of the error covariance decomposition
that can be applied to any kinds of antenna configurations, and then establish the DRT performance as a closed-form.
Our analysis complements the earlier work on DMT \cite{Changick:12JSAC} which is only valid for a positive multiplexing gain,
and thus allows us to fully characterize the diversity order of the MMSE transceivers in MIMO AF relaying systems.
Again, we note that the result of our DRT analysis is unpredictable via DMT analysis.
Finally, simulations results will demonstrate the accuracy of the analysis.


Throughout this paper, normal letters represent scalar quantities,
boldface letters indicate vectors and boldface uppercase letters
designate matrices. We use $\mathbb{C}$ to denote a set of complex numbers.
The superscript $(\cdot)^{H}$ stands for conjugate transpose.
$\Iv_N$ is defined as an $N \times N$ identity matrix, and $E[\cdot]$ and $\lceil\cdot\rceil$ means the expectation
and rounding up to the next higher interger, respectively.
$[\Av]_{k,k}$ and $\textrm{Tr}\left(\Av\right)$ denote the $k$-th diagonal element and trace function of a matrix $\Av$, respectively.
The $k$-th element of a vector $\av$ is denoted by $a_k$.

\section{System Model}\label{sec:System Model}

\begin{figure}
\begin{center}
\includegraphics[width=3.4in]{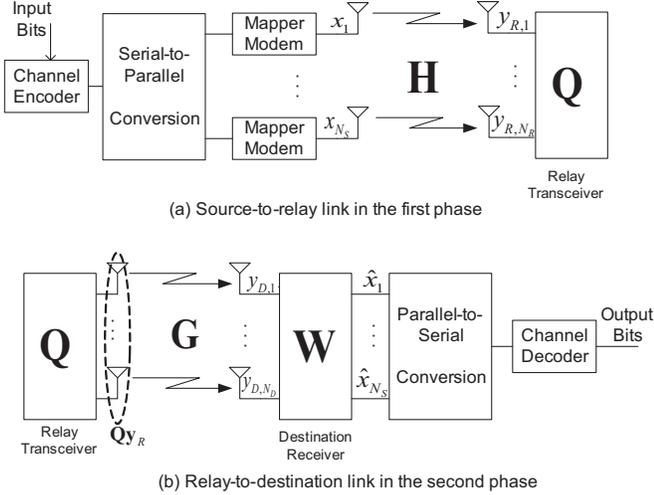}
\end{center}
\caption{Joint encoding/decoding structure for MIMO AF relaying systems with linear MMSE transceivers \label{figure:SystemModel2.eps}}
\end{figure}

Figure \ref{figure:SystemModel2.eps} shows the input-output system model for quasi-static flat fading MIMO AF relaying channels equipped with
$N_S$, $N_R$, and $N_D$ number of antennas at the source, the relay, and the destination, respectively.
A single channel encoder supports all the data streams at the source so that coding is applied jointly across antennas.
We assume no channel state information (CSI) at the source, while both the relay and the destination have perfect CSI of both links.
Due to loop interference at the relay,
it is assumed that each data transmission occurs in two separate phases (time or frequency).
A direct link between the source and the destination is ignored due to large pathloss.

In the first phase, the source transmits the signal vector $\xv=[x_1,x_2,\ldots,x_{N_S}]^T\in\mathbb{C}^{N_S\times1}$ to the relay,
and then the received signals at the relay, $\yv_{R}\in\mathbb{C}^{N_R\times 1}$ is given by
$\yv_{R}=\Hv\xv+\nv_{R}$,
where $\Hv\in\mathbb{C}^{N_R\times N_S}$ and $\nv_{R}\in\mathbb{C}^{N_R\times 1}$ denote the source-to-relay channel matrix and
the noise vector at the relay, respectively.
Due to no CSI at the source, transmit antennas operate with equal power as $E[| x_k|^2]=\frac{P_S}{N_S}=\rho$ for all $k$, where
$P_S$ represents the total transmit power at the source.

In the second phase, the relay signal $\yv_R$ is amplified
by the relay matrix $\Qv\in\mathbb{C}^{N_R\times N_R}$ and transmitted to the destination.
Then, the received signal at the destination is written by
\bea\label{eq:yD}
\yv_{D}=\Gv\Qv\yv_R+\nv_D=\Gv\Qv\Hv\xv+\Gv\Qv\nv_R+\nv_{D},
\eea
where $\nv_{D}$ designates the noise vector at the destination.
Note that the relay matrix $\Qv$ must
satisfy the relay power constraint $P_R$ as $E[\Vert\Qv\yv_R\Vert^2]=\text{Tr}(\Qv(\rho\Hv\Hv^H+\Iv_{N_R})\Qv^H)\leq P_R$.
Finally, when a linear MMSE receiver $\Wv\in\mathbb{C}^{N_S\times N_D}$ is employed at the destination,
the estimated signal waveform $\hat{\xv}\in\mathbb{C}^{N_S\times 1}$ is expressed as $\hat{\xv}=\Wv\yv_D$.

Unlike the open-loop P2P MIMO systems, the diversity order may vary according to the forwarding scheme $\Qv$ at the relay.
In this paper, we examine the diversity order of the MMSE transceivers $\Qv$ and $\Wv$ which are
jointly optimized with respect to the MMSE \cite{Guan:08} \cite{Changick:09TWC}\footnote{The diversity order
of other relaying strategies are currently under investigation for our future works.}.
Throughout the paper, we assume that all channel matrices have random entries which are independent and identically distributed (i.i.d.)
complex Gaussian $\sim\mathcal{CN}(0,1)$, but remain constant over a codeword duration.
All elements of the noise vectors $\nv_R$ and $\nv_{D}$ are also assumed to be i.i.d. $\sim\mathcal{CN}(0,1)$.

\section{Optimal Transceiver Design}\label{sec:Optimal Transceiver Design}

We would like to mention that the MMSE transceiver design between the relay and the destination
has first developed in \cite{Guan:08}.
However, it turns out that the approach in \cite{Guan:08} which is based on the singular-value decomposition
is cumbersome to be dealt with
due to the complicated structure of a compound channel matrix and colored noise at the destination.
In this section, we introduce an alternative design method based on the error covariance decomposition, which makes the analysis more tractable.
This is an extension of the result in \cite{Changick:09TWC}.

We define error vector $\ev\triangleq\hat{\xv}-\xv$ and its covariance matrix $\Rv_e\triangleq E[\ev\ev^H]$.
Then, the joint MMSE optimization problem for $\Qv$ and $\Wv$ is written by
\bea\label{eq:optimization problem}
\min_{\mathbf{Q,W}}\text{Tr}\left(\Rv_e\right)~~s.t.~~\text{Tr}\left(\Qv(\rho\Hv\Hv^H+\Iv_{N_R})\Qv^H\right)\leq P_R.
\eea
By the orthogonality principle $E[\ev\yv_D^H]=\mathbf{0}$, it is easy to find that the optimal receiver at the destination
is given by
$\hat{\Wv}=\rho\Hv^H\Qv^H\Gv^H(\rho\Gv\Qv\Hv\Hv^H\Qv^H\Gv^H+\Iv_{N_D})^{-1}$.
Therefore, the remaining work is now to determine the relay transceiver $\Qv$.

The following lemma \cite[Lemma 1]{SJang:10} shows that
the optimal relay matrix $\Qv$ can be expressed as a product of two matrices.
\begin{Lemma}\label{Lemma:Lemma1}
{\it Under the MMSE strategy, the optimal relay matrix $\Qv$ consists of the relay precoder
$\Bv\in\mathbb{C}^{N_R\times N_S}$ and the relay receiver $\Lv\in\mathbb{C}^{N_S\times N_R}$ as
\bea\label{eq:lemma1}
\hat{\Qv}=\Bv\Lv,
\eea where
$\Bv$ is an arbitrary matrix, while $\Lv=\rho\Hv^H(\rho\Hv\Hv^H+\Iv_{N_R})^{-1}$
is an MMSE receiver for the first hop channel $\Hv$ with input signal $\xv$.}
\end{Lemma}

Now, let us define $\yv\in\mathbb{C}^{N_S\times1}$ as the relay receiver output signal, i.e., $\yv\triangleq\Lv\yv_R$
and its covariance matrix $\Rv_y\triangleq E[\yv\yv^H]\in\mathbb{C}^{N_S\times N_S}$ as
\bea\label{eq:Ry}
\Rv_y=\Lv(\rho\Hv\Hv^H+\Iv_{N_R})\Lv^H.
\eea
Then, the estimated signal vector $\hat{\xv}$ and the relay power constraint in (\ref{eq:optimization problem}) are respectively rephrased as
\bea\label{eq:relay power constraint}
\hat{\xv}=\Wv(\Gv\Bv\yv+\nv_D)~~\text{and}~~\text{Tr}(\Bv\Rv_y\Bv^H)\leq P_R.
\eea
Since the rank of $\Rv_y$ equals $M\triangleq\min(N_S,N_R)$, $\Rv_y$ becomes clearly non-invertible when $N_S>N_R$.
This fact makes the problem more challenging, but has not been fully addressed in conventional literature.
In the following, we revisit the previous works in \cite{Guan:08} and \cite{Changick:09TWC}, and
provide a more generalized and insightful design strategy without restriction on the number of antennas at the source.

In fact, when the relay matrix has the form of (\ref{eq:lemma1}),
the error covariance matrix $\Rv_e$ in (\ref{eq:optimization problem}) can be expressed as a sum of
two individual covariance matrices, each of which represents the first hop and the second hop MIMO channels, respectively.
This result has been proved in \cite{Changick:09TWC}, but the proof was limited to the cases of $N_S\leq\min(N_R,N_D)$.
For the sake of completeness, we give a new result of error decomposition that can be applied to any kind of antenna configurations.

\begin{Lemma}\label{Lemma:Lemma2}
{\it Define the eigenvalue decomposition
$\Rv_y=\Uv_y\mathbf{\Lambda}_{y}\Uv_y^H$ where $\Uv_y\in\mathbb{C}^{N_S\times N_S}$ is a unitary matrix and
$\mathbf{\Lambda}_{y}\in\mathbb{C}^{N_S\times N_S}$ represents a square diagonal
matrix with eigenvalues $\lambda_{y,k}$ for $k=1,\ldots,N_S$ arranged in descending order.
Then, without loss of MMSE optimality, we have
\bea\label{eq:lemma2}
\Rv_e=(\Hv^H\Hv+\rho^{-1}\Iv_{N_S})^{-1}~~~~~~~~~~~~~~~~~~~~~~~~~\nonumber\\
+\widetilde{\Uv}_y\big(\widetilde{\Uv}_{y}^H\Bv^H\Gv^H\Gv\Bv\widetilde{\Uv}_{y}+\widetilde{\mathbf{\Lambda}}_y^{-1}\big)^{-1}\widetilde{\Uv}_{y}^H,
\eea
where $\widetilde{\Uv}_y\in\mathbb{C}^{N_S\times M}$ is a matrix constructed by the first $M$ columns of $\Uv_y$ and
$\widetilde{\mathbf{\Lambda}}_y=\widetilde{\Uv}_y^H\Rv_y\widetilde{\Uv}_y$ indicates
the $M\times M$ upper-left submatrix of $\mathbf{\Lambda}_{y}$.}
\end{Lemma}
\begin{IEEEproof}
As the relay receiver $\Lv$ follows the receive Wiener filter structure,
its output signal $\yv$ should satisfy the orthogonality principle \cite{Joham:05}, i.e., $E\left[\big(\yv-\xv\big)\yv^H\right]=\mathbf{0}$.
Now, using $\yv$, we can express the MSE as
$E\left[\|\ev\|^2\right]=E\left[\left\|\hat{\xv}-\yv+\yv-\xv\right\|^2\right]$.
Then, due to the orthogonality principle above, it is true that
the signal $\yv-\xv$ becomes orthogonal to $\hat{\xv}$ as well as $\yv$, since $\hat{\xv}=\Wv\yv_D=\Wv(\Gv\Bv\yv+\nv_D)$
is also a function of $\yv$ and independent noise $\nv_D$.
Therefore, it follows
\bea\label{eq:two MSE}
E\left[\|\ev\|^2\right]=\text{MSE}_{H}+\text{MSE}_{G}, \nonumber
\eea
where $\text{MSE}_{H}\triangleq E\left[\left\|\yv-\xv\right\|^2\right]$ and $\text{MSE}_{G}\triangleq E\left[\left\|\Wv\yv_D-\yv\right\|^2\right]$.
This result also illustrates that for a given structure of $\Qv=\Bv\Lv$,
the optimal destination receiver $\Wv$ can be alternatively expressed as
$\hat{\Wv}=\Rv_y\Bv^H\Gv^H(\Gv\Bv\Rv_y\Bv^H\Gv^H+\Iv_{N_D})^{-1}$ which amounts to
an MMSE receiver for the second hop channel $\Gv$ with input signal $\yv$.
In what follows, we will show that $\text{MSE}_H$ and $\text{MSE}_G$ in (\ref{eq:two MSE}) can be expressed
as the first and second term in (\ref{eq:lemma2}), respectively.

Let us first have a look at $\text{MSE}_G$. Then, it follows
\bea
\text{MSE}_\text{G}=E[\text{Tr}\left((\Wv\yv_D-\yv)(\Wv\yv_D-\yv)^H\right)]~~~~~~~~~~~~~~~~~~~~~\nonumber\\
=\text{Tr}\big(\Rv_y\!\!-\!\!\Rv_y\Bv^H\Gv^H(\Gv\Bv\Rv_y\Bv^H\Gv^H\!\!\!+\!\Iv_{N_D})^{-1}\Gv^H\Bv^H\Rv_y\big).~~\nonumber
\eea
Now, we write the relay precoder $\Bv$ in a more general form as $\Bv=\breve{\Bv}\Uv_{y}^H$ where $\breve{\Bv}=[\Bv_1~\Bv_2]$
with $\Bv_1\in\mathbb{C}^{N_R\times M}$ and $\Bv_2\in\mathbb{C}^{N_R\times (N_S-M)}$.
Since $\Rv_y$ is a rank $M$ matrix,
setting $\Bv_2=\mathbf{0}$ has no impact on both the MSE and the relay power consumption, i.e., $\text{Tr}(\Bv\Rv_y\Bv^H)$.
Therefore, without loss of generality, $\text{MSE}_\text{G}$ is further rephrased as
\bea
\text{MSE}_\text{G}=\text{Tr}\big(\widetilde{\Uv}_y(\widetilde{\mathbf{\Lambda}}_y-\widetilde{\mathbf{\Lambda}}_y^H\Bv_1^H\Gv^H~~~~~~~~~~~~~~~~~~\nonumber\\
\times(\Gv\Bv_1\widetilde{\mathbf{\Lambda}}_y
\Bv_1^H\Gv^H+\Iv_{N_D})^{-1}\Gv^H\Bv_1^H\widetilde{\mathbf{\Lambda}}_y)\widetilde{\Uv}_y^H\big)\nonumber\\
=\widetilde{\Uv}_y\big(\Bv_1^H\Gv^H\Gv\Bv_1\widetilde{\Uv}_{y}+\widetilde{\mathbf{\Lambda}}_y^{-1}\big)^{-1}\widetilde{\Uv}_{y}^H~~~~~\nonumber\\
=\widetilde{\Uv}_y\big(\widetilde{\Uv}_{y}^H\Bv^H\Gv^H\Gv\Bv\widetilde{\Uv}_{y}+\widetilde{\mathbf{\Lambda}}_y^{-1}\big)^{-1}\widetilde{\Uv}_{y}^H,~\nonumber
\eea
where the last equality follows from $\Bv_1=\Bv\widetilde{\Uv}_y$.

Meanwhile, the case of $\text{MSE}_H$ is equivalent to a situation of P2P MIMO channels with the input signal vector $\xv$.
Thus, the proof simply follows the previous results in \cite{Joham:05}, and thus omitted.
\end{IEEEproof}

When $N_S\leq N_R$, Lemma \ref{Lemma:Lemma2} is equivalent to one in \cite{Changick:09TWC}; thus is more general.
Now, the result of Lemma \ref{Lemma:Lemma2} illustrates that
the original joint optimization problem in (\ref{eq:optimization problem}) can be reduced to optimizing $\Bv$,
since the first term of $\Rv_e$ consists of known parameters.
Define eigenvalue decomposition $\Gv^H\Gv=\Vv_g\mathbf{\Lambda}_g\Vv_g^H$
where $\mathbf{\Lambda}_g$ designates a square diagonal
matrix with eigenvalues $\lambda_{g,k}$ for $k=1,\ldots,N_d$ arranged in descending order.
Then, we can show that the optimal relay precoder $\Bv$ can be generally written by
$\hat{\Bv}=\widetilde{\Vv}_g\mathbf{\Phi}\widetilde{\Uv}_y^H$ where $\widetilde{\Vv}_g\in\mathbb{C}^{N_R\times M}$ denotes a matrix constructed by the first $M$ columns
of $\Vv_g$ and $\mathbf{\Phi}\in\mathbb{C}^{M\times M}$ is an arbitrary matrix \cite{Guan:08}.

Now, substituting $\hat{\Bv}$ into (\ref{eq:lemma2}), the modified problem determines the optimal $\mathbf{\Phi}$:
\bea\label{eq:modified problem}
\hat{\mathbf{\Phi}}=\arg\min_{\mathbf{\Phi}}~\big(\mathbf{\Phi}\widetilde{\mathbf{\Lambda}}_g\mathbf{\Phi}^H+\widetilde{\mathbf{\Lambda}}_y^{-1}\big)^{-1}~\textit{s.t.}~
\text{Tr}(\mathbf{\Phi}\widetilde{\mathbf{\Lambda}}_y\mathbf{\Phi}^H)\leq P_R.\nonumber
\eea
Here $\widetilde{\mathbf{\Lambda}}_g$ represents
the $M\times M$ upper-left submatrix of $\mathbf{\Lambda}_g$.
Since we have $\text{Tr}(\Av^{-1})\geq \sum_{i=1}^{M}\left([\Av]_{k,k}\right)^{-1}$ for a positive definite matrix $\Av$ \cite{Komaroff:90},
it is easy to check that the minimum MSE is achieved when $\mathbf{\Phi}$ is a diagonal matrix,
which leads to a simple convex problem.
The remaining procedure simply follows from previous works in \cite{Guan:08}, \cite{Changick:09TWC}, and \cite{Palomar:03}.
Finally, in combination with the relay receiver $\Lv$ in (\ref{eq:lemma1}), we have
\bea\label{eq:optimal relay matrix}
\hat{\Qv}=\hat{\Bv}\Lv=\widetilde{\Vv}_g\hat{\mathbf{\Phi}}\widetilde{\Uv}_y^H\Lv,
\eea
where the $k$-th diagonal element of $\hat{\mathbf{\Phi}}$ is determined by
$|\hat{\phi}_k|^2=\frac{1}{\lambda_{y,k}\lambda_{g,k}}\left(\sqrt{\frac{\lambda_{y,k}\lambda_{g,k}}{\nu}}-1\right)^+$ for $k=1,2,\ldots,M$ with
$(\cdot)^+=\max(\cdot,0)$ and $\nu$ being chosen to satisfy the relay power constraint in (\ref{eq:relay power constraint}).
Note that if $\lambda_{g,k}=0$, we have $|\hat{\phi_k}|^2=0$.

\section{Diversity-Rate Tradeoff Analysis}\label{sec:DRT Analysis}

We now investigate the diversity order of the MMSE optimal transceiving scheme in
MIMO AF relaying systems studied in the previous section, where
data streams are jointly encoded across the antennas at the source (vertical encoding).
The diversity analysis may be conducted by either outage probability or
pairwise error probability (PEP) \cite{Mehana:12}.
In this paper, we focus on the outage probability of mutual information (MI) assuming infinite length Gaussian codewords.
For simplicity, we assume that $P_R=P_T=\rho N_t$, but the result can be easily extended to more general cases.
We say that two functions $f(\rho)$ and $g(\rho)$ are exponentially equal when
\bea
\lim_{\rho\rightarrow\infty}\frac{\log f(\rho)}{\log \rho}=\lim_{\rho\rightarrow\infty}\frac{\log g(\rho)}{\log \rho},\nonumber
\eea
and denoted by $f(\rho)\doteq g(\rho)$.

When the coding is applied across antennas with MMSE receivers, the MI is defined as \cite{Hedayat:07}
\bea\label{eq:MI def}
\mathcal{I}=\frac{1}{2}\sum_{k=1}^{N_S}\log\left(1+\gamma_k\right),\nonumber
\eea
where $\gamma_k=\rho/[\Rv_e]_{k,k}-1$.
Then, we obtain
\bea\label{eq:MI lowerbound}
\mathcal{I}\overset{(a)}{\geq}-\frac{N_S}{2}\log\Big(\frac{1}{\rho N_S}\text{Tr}(\Rv_e)\Big)~~~~~~~~~~~~~~~~~~~~~\nonumber\\
\overset{(b)}{=}-\frac{N_S}{2}\log\Big(\frac{1}{\rho N_S}\big(\text{Tr}\big(\Hv^H\Hv+\rho^{-1}\Iv_{N_S}\big)^{-1}~~\nonumber\\
+ \text{Tr}\big(\hat{\mathbf{\Phi}}^H\widetilde{\mathbf{\Lambda}}_g\hat{\mathbf{\Phi}}+\widetilde{\mathbf{\Lambda}}_{y}^{-1}\big)^{-1}\big)\Big)\nonumber\\
\overset{(c)}{\geq}-\frac{N_S}{2}\log\Big(\frac{1}{N_S}\big(\text{Tr}(\rho\mathbf{\Lambda}_h+\Iv_{N_S})^{-1}~~~~~~~~~~~\nonumber\\
+\text{Tr}\big(\eta\rho\widetilde{\mathbf{\Lambda}}_g+\rho\widetilde{\mathbf{\Lambda}}_{y}^{-1}\big)^{-1}\big)\Big),
\eea
where (a) follows from the Jensen's inequality, (b) is due to the optimal relay precoder $\hat{\Bv}$ described in (\ref{eq:optimal relay matrix}),
and (c) holds by setting $\hat{\mathbf{\Phi}}=\sqrt{\eta}\Iv_{N_M}$,
where $\eta$ can be chosen to be $1$ to satisfy the relay power constraint in (\ref{eq:relay power constraint}) (see Appendix \ref{Appendix:Appendix 2}).
Let us define the outage probability as $P_{\text{out}}\triangleq\left(\mathcal{I}\leq R\right)$.
Then, using the MI bound in (\ref{eq:MI lowerbound}) and setting the target data rate as $R$,
we obtain the outage probability upperbound as $P_{\text{out}}\leq P_{\text{out}}^U$, where
\bea\label{eq:outage definition}
P_{\text{out}}^U \triangleq P\bigg(\sum_{k=1}^{M}\frac{1}{1+\rho\lambda_{h,k}}+\sum_{k=1}^{M}\frac{1}{\rho\lambda_{g,k}+\rho\lambda_{y,k}^{-1}}\geq m\bigg),
\eea
with $m\triangleq N_S 2^{-\frac{2R}{N_S}}-(N_S-M)$.

First, let us first set the target data rate as $R=r\log\rho$.
Then, the resulting outage exponent leads to the DMT performance which captures
the tradeoff between the multiplexing gain $r$ and block error probability at high SNR ($\rho\rightarrow\infty$).
\begin{Theorem}\label{Theorem:Theorem1}
{\it For MIMO AF relaying systems with positive multiplexing gain $r>0$, the achievable DMT of the MMSE transceivers is given by}
\bea\label{eq:Theorem1}
d(r)=\left\{\begin{array}{cc}\!\!\!\!(N_R-N_S+1)\left(1-\frac{2r}{N_S}\right)^+ & \!\!\!\!\!\text{if}~ N_S\leq \min(N_R,N_D) \\
0 & \text{otherwise}\end{array}\right.\nonumber
\eea
\end{Theorem}
\begin{IEEEproof}
The proof is simply obtained from \cite{Changick:12JSAC}
by assuming that the direct link between the source and the destination can be ignored.
Details are omitted for brevity.
\end{IEEEproof}

As described in Theorem \ref{Theorem:Theorem1},
the DMT analysis accurately predicts the diversity order of
the MMSE transceivers when the multiplexing gain is positive ($r>0$).
However, when the target rate $R$ is fixed with respect to $\rho$, i.e., $r=0$ and sufficiently low,
it is observed that the performance is in stark contrast to one predicted by the DMT analysis.
In the following, we will analyze the fixed rate diversity of the MMSE transceivers
as a function of rate $R$ and the number of antennas at each node.

\begin{Theorem}\label{Theorem:Theorem2}
{\it For MIMO AF relaying systems with fixed rate $R$ ($r=0$), the achievable DRT of the MMSE transceivers is
\bea\label{eq:Theorem2}
d(R)=\min\Big(\overline{m}(N_R+N_S-2M+\overline{m}),~~~~~~~~~~~~~~\nonumber\\
(N_R-M+\overline{m})(N_D-M+\overline{m})^+\Big),\nonumber
\eea
where $\overline{m}\triangleq \left\lceil \big( N_S 2^{-\frac{2R}{N_S}}+M-N_S\big)^+ \right\rceil$.}
\end{Theorem}
\begin{IEEEproof}
We begin by defining $\alpha_k\triangleq-\log\lambda_{h,k}/\log\rho$ and $\beta_k\triangleq-\log\lambda_{g,k}/\log\rho$ for
$k=1,\ldots,M$. Then, $P_{\text{out}}^U$ in (\ref{eq:outage definition}) is alternatively expressed as
\bea\label{eq:outage upperbound}
P_{\text{out}}^U~~~~~~~~~~~~~~~~~~~~~~~~~~~~~~~~~~~~~~~~~~~~~~~~~~~~~~~~~~~~~~\nonumber\\
\!\!\overset{(a)}{=}\!\!P\Big(\sum_{k=1}^{M}\frac{1}{1+\rho^{1-\alpha_k}}+\sum_{k=1}^{M}\frac{1}{1+\rho^{1-\beta_k}+\rho^{-(1-\alpha_k)}}\geq m\Big)\nonumber\\
\!\!\overset{(b)}{\doteq} \!\!P\Big(\sum_{k=1}^{M}\frac{1}{1+\rho^{1-\alpha_k}}+\sum_{k=1}^{M}\frac{1}{1+\rho^{1-\beta_k}+\rho^{-(1-\alpha_k)}}\geq \overline{m}\Big)
\eea
where (a) follows from $\rho\lambda_{y,k}^{-1}=1+\rho^{-(1-\alpha_k)}$ (see the definition of $\Rv_y$ in (\ref{eq:Ry}) and (\ref{eq:Ry2}))
and (b) is due to the fact that if $m<0$, the outage always occurs.
Now, at high SNR, we can write the exponential equality as
\bea\label{eq:alpha_k}
\frac{1}{1+\rho^{1-\alpha_k}}\doteq\left\{\begin{array}{cc} 0 & \text{if}~~\alpha_k<1 \\
1 & \text{if}~~\alpha_k>1 \end{array}\right.~~~~~~~~~~~~~~~\\
\label{eq:beta_k}
\frac{1}{1+\rho^{1-\beta_k}+\rho^{-(1-\alpha_k)}}\doteq\left\{\begin{array}{cc} 0 & \!\!\!\text{if}~~\alpha_k>1~\text{or}~\beta_k<1 \\
1 & \text{if}~~\alpha_k<1~\text{and}~\beta_k>1 \end{array}\right.,
\eea
for all $k=1,\ldots,M$. Note that in an asymptotic sense $\rho\rightarrow\infty$, the cases where the eigenvalues
take on values that are comparable with $1/\rho$ can be ignored \cite{Kumar:09}.

We first see from these results that in order for the outage to occur,
at least $\overline{m}$ number of terms should be $1$ among $2M$ summation terms in (\ref{eq:outage upperbound}).
The above results also reveal that two terms in (\ref{eq:alpha_k}) and (\ref{eq:beta_k}) cannot be simultaneously $1$ at a certain $k$.
As will be clear later in this proof,
this property allows us to obtain the full diversity order as $\overline{m}$ tends to be large.
Remind that all eigenvalues are in descending order, which means that
$\{\alpha_i\}$ and $\{\beta_i\}$ are ordered according to $\alpha_1\leq\cdots\leq\alpha_M$ and $\beta_1\leq\cdots\leq\beta_M$.
Thus, if $\alpha_1>1$, the term in (\ref{eq:beta_k}) converge to zero for all $k$, regardless of $\beta$.

For all $i=1,\ldots,M$, let us define all possible events in which $i$ number of terms in (\ref{eq:outage upperbound}) equal $1$ as
\bea
\mathcal{E}_{h,i}\triangleq\!\big\{\alpha_{M-i+1}>1>\alpha_{M-i}\big\}~~\text{and}~~~~~~~~~~~~~~~~~~~~~~~~~~~~\nonumber\\
\mathcal{E}_{g,i,j}\triangleq\!\big\{\beta_{M-i+1}>1>\beta_{M-i}\big\}\cap\mathcal{E}_{h,j}~\text{for}~j=0,1,\ldots,i\!-\!1,\nonumber
\eea
Then, it follows from the union bound that
\bea\label{eq:Union to summation}
P_{\text{out}}^U&=&P\Big(\bigcup_{i=\overline{m}}^M\Big[\mathcal{E}_{h,i}\cup\Big(\bigcup_{j=0}^{i-1}\mathcal{E}_{g,i,j}\Big)\Big]\Big)\nonumber\\
&\leq&\sum_{i=\overline{m}}^M\Big(P(\mathcal{E}_{h,i})+\sum_{j=0}^{i-1}P(\mathcal{E}_{g,i,j})\Big),
\eea

First, we define $P(\mathcal{E}_{h,i})\doteq\rho^{-d_{h,i}(R)}$, $i=1,\ldots,M$. Then,
applying Varadhan's lemma as in \cite{Kumar:09} by using the pdf\footnote{The pdf is slightly different from \cite{Kumar:09}, since the eigenvalue ordering is reversed.}
of the random vector $\av=[\alpha_1,\ldots,\alpha_M]$ as
\bea
f(\av)
\doteq\Big[\prod_{l=1}^M\rho^{-(N_S+N_R-2l+1)\alpha_l}\Big]\text{exp}\Big(-\sum_{l=1}^M\rho^{-\alpha_l}\Big),\nonumber
\eea
we obtain
\bea\label{eq:d_ih(R)}
d_{h,i}(R)&=&\inf_{\av\in\mathcal{E}_{h,i},\forall\alpha_l>0}\sum_{l=1}^{M}(N_S+N_R-2l+1)\alpha_l\nonumber\\
&=&\sum_{l=1}^{M-i}(N_S+N_R-2l+1)\times0\nonumber\\
&&+\sum_{l=M-i+1}^{M}(N_S+N_R-2l+1)\times1\nonumber\\
&=&i(N_R+N_S-2M+i).
\eea

Now, let us examine the probability of the event $\mathcal{E}_{g,i,j}$, i.e., $P(\mathcal{E}_{g,i,j})\doteq\rho^{-d_{g,i,j}(R)}$.
Defining $L\triangleq\min(N_R,N_D)$, the pdf of the random vector $\bv=[\beta_1,\ldots,\beta_L]$ is given by
\bea
f(\bv)\doteq\Big[\prod_{l=1}^L\rho^{-(N_R+N_D-2l+1)\beta_i}\Big]\text{exp}\Big(-\sum_{l=1}^L\rho^{-\beta_l}\Big).\nonumber
\eea
Then, the probability of the event $\mathcal{E}_{g,i,j}$ is
\bea
P(\mathcal{E}_{g,i,j})=\int_{\mathcal{E}_{g,i,j}}f(\av,\bv)d\av d\bv~~~~~~~~~~~~~~~~~~~~~~~~~~~~~~~~\nonumber\\
\doteq\int_{\mathcal{E}_{g,i,j}}\left[\rho^{-\sum_{l=1}^M (N_S+N_R-2l+1)\alpha_l-\sum_{l=1}^L (N_R+N_D-2l+1)\beta_l}\right]\nonumber\\
\times\text{exp}\Big(-\sum_{l=1}^M\rho^{-\alpha_l}-\sum_{l=1}^L\rho^{-\beta_l}\Big)d\av d\bv,~~~~~~~~\nonumber
\eea
due to the independence of $\av$ and $\bv$, and applying Varadhan's lemma again, we have
\bea\label{eq:d_ig(R)}
\!\!\!\!d_{g,i,j}(R)=\inf_{\small \begin{array}{c}(\av,\bv)\in\mathcal{E}_{g,i,j},\\ \forall\alpha_l,\forall\beta_l>0\end{array}}
\sum_{l=1}^{M}(N_S+N_R-2l+1)\alpha_l\!\!\nonumber\\
+\sum_{l=1}^{L}(N_R+N_D-2l+1)\beta_l~~~~~~~~~~~~\nonumber\\
=\sum_{l=M-j+1}^{M}(N_S+N_R-2l+1)~~~~~~~~~~~~~~~~~~~~~~~~\nonumber\\
+\sum_{l=M-i+1}^{L}(N_R+N_D-2l+1)~~~~~~~~~~~\nonumber\\
=j(N_S+N_R-2M+j)~~~~~~~~~~~~~~~~~~~~~~~~~~~~~~~\nonumber\\
+(N_R+N_D-L-M+i)(L-M+i)^+~~~~~~\nonumber\\
=j(N_S+N_R-2M+j)~~~~~~~~~~~~~~~~~~~~~~~~~~~~~~~\nonumber\\
+(N_R-M+i)(N_D-M+i)^+.~~~~~~~~~~~
\eea

From the results in (\ref{eq:Union to summation}-\ref{eq:d_ig(R)}), we eventually conclude that
\bea
P_{\text{out}}~\dot{\leq}~ P(\mathcal{E}_{h,\overline{m}})+P(\mathcal{E}_{g,\overline{m},0})\doteq\rho^{-\min(d_{h,\overline{m}}(R),d_{g,\overline{m},0}(R))},\nonumber
\eea
because all other events causing the outage in (\ref{eq:Union to summation}) yield higher outage exponents
than $\mathcal{E}_{h,\overline{m}}$ or $\mathcal{E}_{g,\overline{m},0}$; thus can be ignored,
and the proof is concluded.
\end{IEEEproof}

Our result in Theorem \ref{Theorem:Theorem2} confirms and complements the earlier work on DMT in Theorem \ref{Theorem:Theorem1}.
We first see that when the rate is high, i.e., $R\geq\frac{N_S}{2}\log N_S$ (or $\overline{m}=1$),
both Theorem \ref{Theorem:Theorem1} and \ref{Theorem:Theorem2} yield the same diversity order.
At high rate, therefore, the diversity order of the MMSE transceivers may be predictable by DMT analysis with setting $r=0$,
and thus very suboptimal compared to the optimal (ML) diversity \cite{DGunduz:10} \cite{Tang:07}.
However, as the rate becomes lower,
it is shown from Theorem \ref{Theorem:Theorem2} that higher diversity order is actually achievable than one predicted by the DMT analysis.
In particular, when $R<\frac{N_S}{2}\log\frac{N_S}{N_S-1}$ (or $\overline{m}=M$),
the MMSE transceivers even exhibit the full diversity order $d=N_R\min(N_S,N_D)$,
thereby achieving an ML-like performance.
It is also interesting to observe that when the rate is sufficiently small,
a certain diversity gain is still achievable even when $N_S>\min(N_R,N_D)$, which is often overlooked in MMSE-based relaying systems.


\section{Numerical Results } \label{sec:Numerical Results}

\begin{figure}
\begin{center}
\includegraphics[width=3.3in]{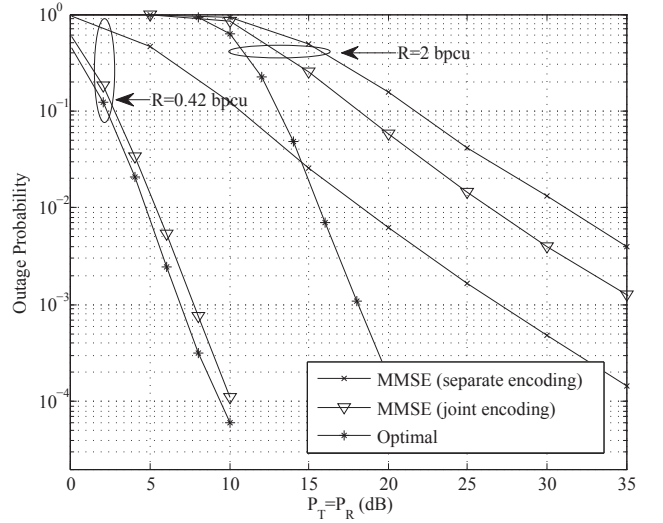}
\end{center}
\caption{Outage probabilities of $2\times2\times2$ MIMO AF relaying systems with R=0.42 and 2 bpcu \label{figure:2x2x2.eps}}
\end{figure}

\begin{figure}
\begin{center}
\includegraphics[width=3.3in]{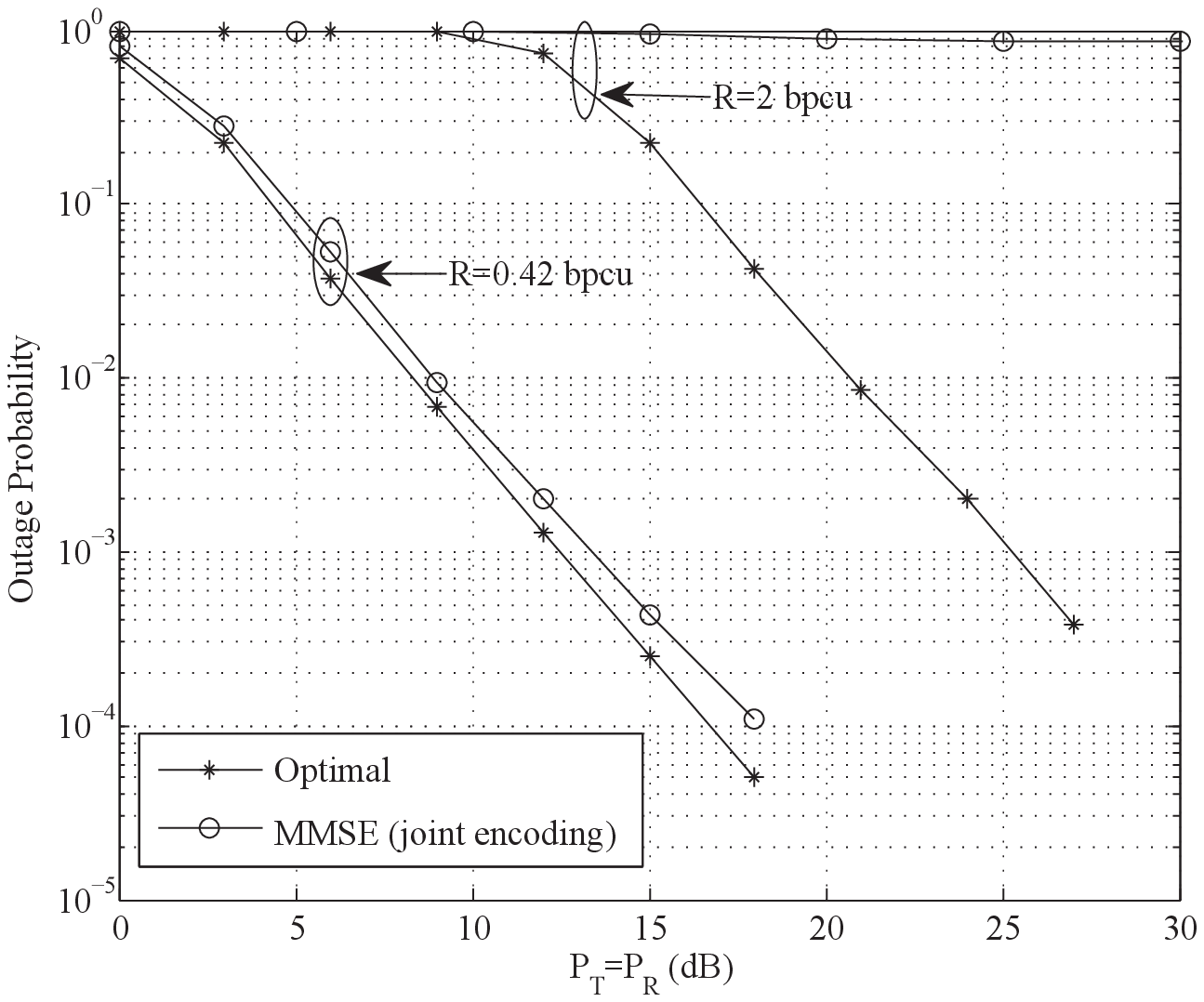}
\end{center}
\caption{Outage probabilities of $2\times2\times1$ MIMO AF relaying systems with R=0.42 and 2 bpcu \label{figure:2x2x1.eps}}
\end{figure}

In this section, we demonstrate the accuracy of our analysis using numerical simulations
for the quasi-static i.i.d. Rayleigh fading model.
Target rate $R$ is measured in bits per channel use (bpcu).
The notation $N_S\times N_R\times N_D$ is used to denote a system with $N_S$ source, $N_R$ relay and $N_D$ destination antennas.
Figure \ref{figure:2x2x2.eps} shows the case of $2\times2\times2$ MIMO AF relaying systems with $R=0.42$ ($\overline{m}=2$) and $2$ ($\overline{m}=1$) bpcu, which
leads to diversity order $4$ and $1$, respectively.
Here, {\it ``Optimal''} indicates the capacity achieving relaying scheme with the optimal receiver (ML) at the destination \cite{Tang:07}.
As predicted by our analysis, it is shown that as the rate becomes smaller, the MMSE transceiver with a joint encoding/decoding structure
as in Figure \ref{figure:SystemModel2.eps}
denoted by {\it "MMSE (joint encoding)"} exhibits near optimal performance
with full diversity order, while the separate encoding gives a constant diversity for all rates.
In Figure \ref{figure:2x2x1.eps}, simulation results for $2\times2\times1$ systems are given.
This result illustrates that even when $N_S>\min(N_R,N_D)$, the MMSE scheme is still able to achieve a certain diversity gain at a low rate.
This observation is flatly conflict with the assumption $N_S\leq\min(N_R,N_D)$ commonly adopted in designs of MMSE-based MIMO AF relaying systems.
Remind that our design method in Section \ref{sec:Optimal Transceiver Design} can be applied to any kinds of antenna configuration
without hurting the MMSE optimality.

\section{Conclusion} \label{sec:Conclusion}

In this paper, we investigated the DRT performance of the linear MMSE transceivers in MIMO AF relaying systems
for all fixed rates and for any number of source, relay, and destination antennas.
First, we generalized the previous error covariance decomposition lemma
so that it can be applied to any kind of antenna configurations.
Then, we derived the achievable DRT as a closed-form, which precisely characterizes
the rate dependent behavior of the MMSE transceivers.
Our analysis allows us to completely characterize the diversity order of the MMSE transceivers together with the DMT
which is only valid for a positive multiplexing gain.
Finally, the analysis was confirmed by numerical simulations.

\appendices

\section{Choosing $\eta$ in (\ref{eq:MI lowerbound})} \label{Appendix:Appendix 2}

From the definition of $\Rv_y$ in (\ref{eq:relay power constraint}), we obtain
\bea\label{eq:Ry2}
\Rv_y=\rho\Hv^H\Hv(\Hv^H\Hv+\rho^{-1}\Iv_{N_S})^{-1}~~~~~~~~~~~~~~~~~~~~~~\nonumber\\
=\rho(\Hv^H\Hv+\rho^{-1}\Iv_{N_S}-\rho^{-1}\Iv_{N_S})(\Hv^H\Hv+\rho^{-1}\Iv_{N_S})^{-1}\nonumber\\
=\rho\Iv_{N_S}-(\Hv^H\Hv+\rho^{-1}\Iv_{N_S})^{-1}.~~~~~~~~~~~~~~~~~~~~~~~~~~
\eea
Then, it is obvious that $\Rv_y\prec\rho\Iv_{N_S}$ where $\prec$ or $\succ$ represent generalized inequality defined on the positive definite cone.
Since we have $\text{Tr}(\Xv)<\text{Tr}(\Yv)$ for positive definite matrices $\Xv\prec\Yv$, it follows that
$\text{Tr}(\Bv\Rv_y\Bv^H)<\text{Tr}(\rho\Bv\Bv^H)=\text{Tr}(\rho\mathbf{\Phi}\mathbf{\Phi}^H)$; thus setting $\mathbf{\Phi}=\Iv_M$ in (\ref{eq:MI lowerbound})
satisfies the relay power constraint (\ref{eq:relay power constraint}).

\bibliographystyle{ieeetr}

\input{bibliography.filelist}


\end{document}